\documentstyle[aps,twocolumn,epsfig,rotating]{revtex}

\begin{document}

\draft \preprint{TRIUMF/McMaster U.}

\begin{title} {\large\bf Static Critical Behavior of the Spin-Freezing
Transition in the Geometrically Frustrated Pyrochlore Antiferromagnet
Y$_2$Mo$_2$O$_7$}

\end{title}


\author{M.J.P. Gingras$^{1,*}$, C.V. Stager$^2$, N.P. Raju$^3$, B.D.
Gaulin$^2$, and J.E. Greedan$^3$}

\address{$^{1}$TRIUMF, Theory Group, 4004 Wesbrook Mall, Vancouver,
B.C., V6T-2A3, Canada} \vspace{-5truemm}

\address{$^{2}$Department of Physics and Astronomy, McMaster
University, Hamilton, ON L8S 4M1, Canada} \vspace{-5truemm}

\address{$^{3}$Department of Chemistry, McMaster University, Hamilton,
ON L8S 4M1, Canada}

\date{\today}

\maketitle

\begin{abstract} \noindent Some frustrated pyrochlore antiferromagnets,
such as Y$_2$Mo$_2$O$_7$, show a spin-freezing transition and magnetic
irreversibilities below a temperature $T_f$ similar to what is observed
in {\it randomly} frustrated spin glasses.  We present results of DC
nonlinear magnetization measurements on Y$_2$Mo$_2$O$_7$ that provide
strong evidence that there is an underlying thermodynamic phase
transition at $T_f$, which is characterized by critical exponents
$\gamma \approx 2.8$ and $\beta \approx 0.8$. These values are typical
of those found in random spin glasses, despite the fact that the level
of random disorder in Y$_2$Mo$_2$O$_7$ is immeasurably small.
\end{abstract}

\vspace{5mm}

\noindent$^*$Present address: Department of Physics,
University of Waterloo, Waterloo,
ON, N2L 3G1, Canada


\narrowtext

\noindent The past five years have seen a resurgence of significant interest
devoted to the systematic study of geometrically frustrated
antiferromagnets~\cite{coey,book,ramirez_review,kawamura}.
Geometric frustration arises in materials containing
antiferromagnetically-coupled magnetic moments which reside on
geometrical units, such as triangles and tetrahedra, that inhibit the
formation of a collinear magnetically-ordered state. The main
motivation  for the current interest in these systems stems from
suggestions that (i) they may display critical phenomena
belonging to a ``new'' chiral universality class different from the
universality classes of collinear magnets~\cite{book,kawamura},
or (ii) the increased
propensity of frustrated antiferromagnets for quantum
zero-temperature spin fluctuations compared to collinear
antiferromagnets might be sufficient to destroy N\'eel order and
drive these systems into novel
non-classical quantum disordered ground states~\cite{book,ramirez_review}.

Systems of classical Heisenberg spins residing on lattices
of corner-sharing triangles or tetrahedra
 and antiferromagnetically coupled via nearest-neighbor
exchange constitute particularly interesting cases of highly frustrated
antiferromagnets (see Fig. 1).  Here, theory~\cite{villain,kag-pyr-theory} and
numerical work~\cite{kag-pyr-num}, show that these systems do not order
and remain in a ``collective paramagnetic
state''~\cite{villain} down to zero temperature.  Since, even for
classical spins, these systems have such a small tendency to
order, they are excellent candidates to display
exotic quantum disordered ground
states~\cite{book,ramirez_review,kagome-S12}. However, and 
perhaps most interestingly,
experiments show that some nominally perfect (i.e.
disorder-free)~\cite{disorder-level} pyrochlore
antiferromagnets~\cite{pyro-sg-behavior} exhibit a spin-freezing
transition at some temperature $T_f$, below which they develop magnetic
irreversibilities (see Fig. 1) and long-time magnetic
relaxation similar to what is found in conventional randomly frustrated
spin glasses such as CuMn, EuSrS, and CdMnTe~\cite{BYFH}.
\begin{figure}
 \begin{center}
    \begin{turn}{90}
        {\epsfig{file=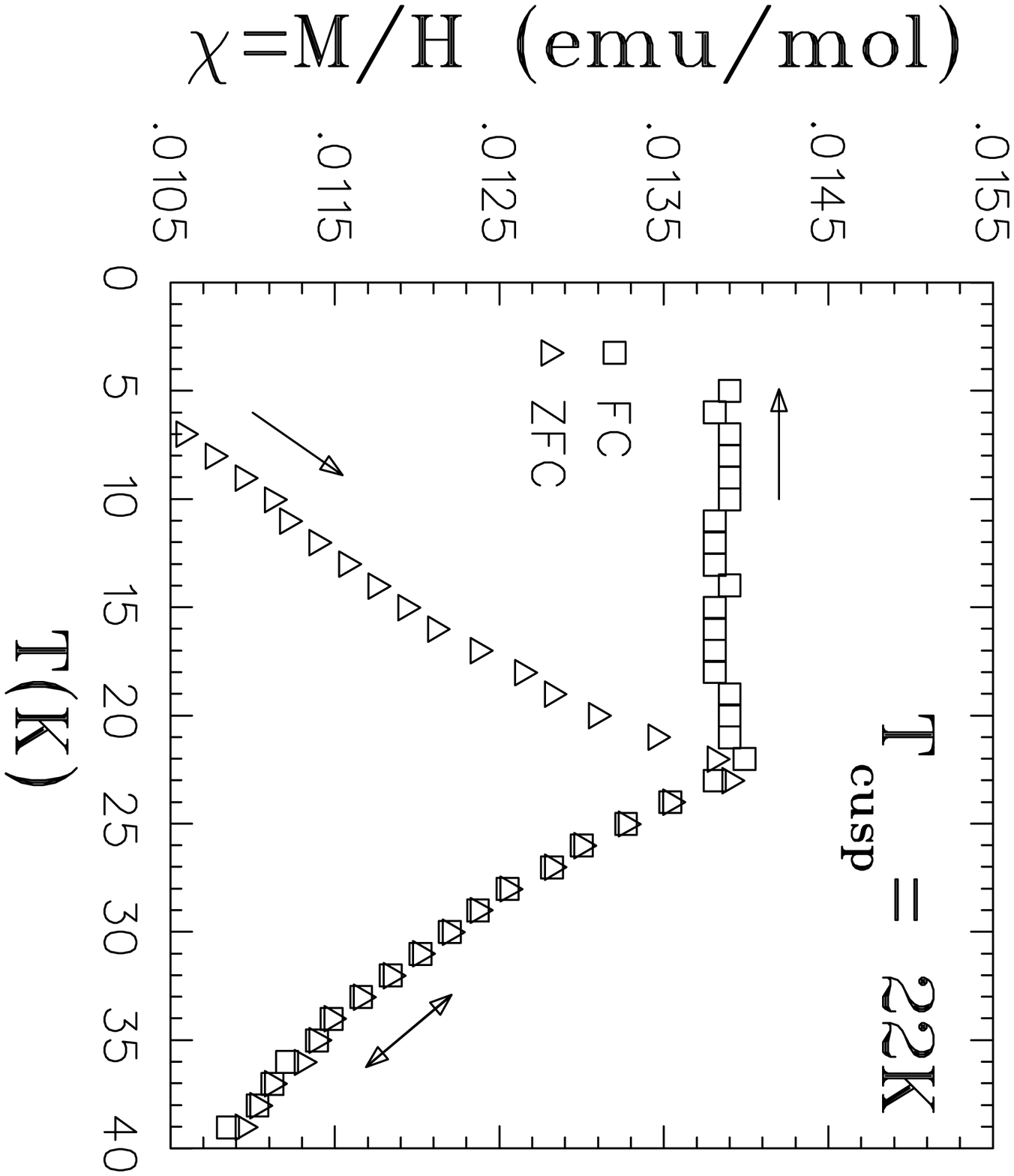,height=5cm,width=5cm} }
        \end{turn}
\end{center}
\end{figure}
\vspace{-2.5cm}
\begin{figure}
 \begin{center}
  {\begin{turn}{0}
        {\epsfig{file=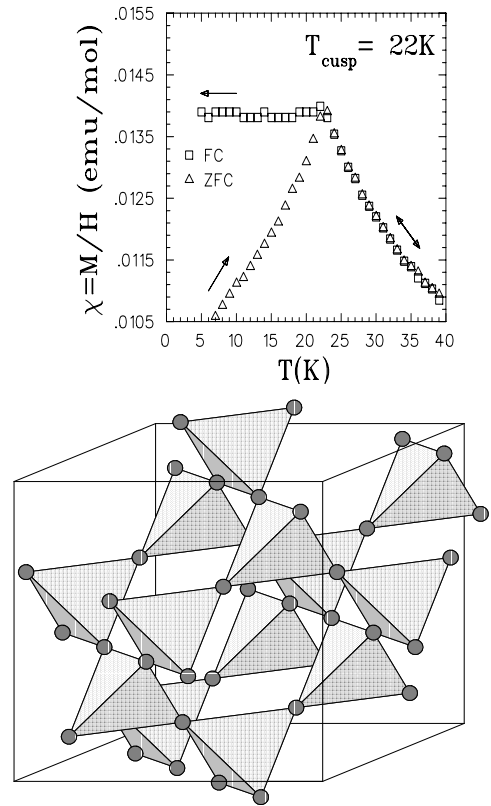,height=11.2cm,width=8cm}  }
        \end{turn} }
\vspace{-3.5cm}
\caption{The top figure shows the field-cooled (FC,  squares)
and zero field-cooled (ZFC,  triangles) magnetization for 
Y$_2$Mo$_2$O$_7$ in a magnetic field
100Oe. The bottom figure shows the
three-dimensional network of corner-sharing tetrahedra
formed by the Mo$^{4+}$ sublattice in Y$_2$Mo$_2$O$_7$.}
\end{center}
\end{figure}
Two important questions arise: Firstly, what is the microscopic origin
of the glassy behavior in pyrochlore antiferromagnets?  Is it due to
the yet undetected microscopic disorder inherent to any real material,
or is it intrinsic to the idealized perfect material?  Secondly,
irrespective of the origin of the glassy behavior, one would like to
know if the spin-freezing is strictly dynamical (i.e. where the system's
relaxation time exceeds the time scale set by the
experimental probe), or is it due to an underlying thermodynamic
transition characterized by a truly divergent (spin-glass) correlation
length and time scale as is believed to occur in
conventional disordered spin
glasses~\cite{BYFH}?  To address these two
questions, we have measured the nonlinear magnetic susceptibility, $\chi_{\rm
nl}$, of the pyrochlore antiferromagnet Y$_2$Mo$_2$O$_7$. 
Briefly, we have found (i) that the
freezing at $T_f$ is well characterized as a {\it thermodynamic} transition
displaying a power-law divergence of the nonlinear
susceptibility coefficient $\chi_3(T) \sim (T-T_f)^{-\gamma}$ with
$\gamma \approx 2.8$,
and that the net nonlinear susceptibility, $\chi_{\rm nl}$, exhibits
critical behavior and temperature-field scaling properties close to
$T_f$ which gives a critical exponent, $\beta \approx
0.8$. The values we obtain for  $\gamma$ and $\beta$ are typical of those found
in conventional disordered spin glasses, despite 
any obvious microscopic disorder in Y$_2$Mo$_2$O$_7$~\cite{disorder-level}.
High temperature measurements of the susceptibility 
 in the Curie-Weiss regime, $\chi =
  C/(T - \theta)$ show that $\theta \sim -200$K, with a Curie constant
  $C$ giving an effective magnetic moment of the order of 2.5 Bohr
  magneton per Mo$^{4+}$~\cite{hubert}. This is very close to the theoretically
  {\it  expected} value of 2.8 Bohr magneton per Mo$^{4+}$.  We have recently
  performed high temperature
  susceptibility measurements in the range 10K$-$800K which are in
  agreement with Hubert's results~\cite{raju_high-T}.
  Clearly, these
  indicate that the susceptibility, and thereby the
  spin-glass-behavior observed in Y$_2$Mo$_2$O$_7$ is not due to a small
  fraction of the spins,  but arises from
  the very large majority of the moments. Essentially, $\chi$
  is small close to $T_f=23$K because $\theta$ is so large and since
  Y$_2$Mo$_2$O$_7$ is an antiferromagnet (i.e $\theta$ is negative).

Y$_2$Mo$_2$O$_7$ is a narrow band gap semiconductor where the Mo$^{4+}$
ions are magnetic, with an antiferromagnetic nearest-neighbor Mo-Mo
superexchange, while Y$^{3+}$ is diamagnetic. Figure 1 shows the cubic
unit cell of the  Mo$^{4+}$ magnetic lattice where there is a magnetic
Mo moment on the vertices of the tetrahedra.  The 270mg powder
sample of Y$_2$Mo$_2$O$_7$ was prepared as described in
Reference~\cite{disorder-level}. Neutron and X-ray powder diffraction
studies show that there is no measurable amount of oxygen vacancies or
intermixing between the Y$^{3+}$ and the Mo$^{4+}$
sublattices~\cite{disorder-level}.  The random disorder in that material,
most likely oxygen vacancies,
is therefore below the 1\% detectability level.  The
magnetization was measured using a commercial SQUID (Quantum Design,
San Diego) magnetometer. The bulk magnetization of Y$_2$Mo$_2$O$_7$
becomes history dependent below $T_f\approx 22$K: the field-cooled (FC) and
zero field-cooled (ZFC) magnetizations measured in a field of 100Oe show
a sharp breakaway below 22K (see Fig.
1)~\cite{disorder-level}.

To determine whether or not a true thermodynamic spin-freezing
transition occurs around $T_f\approx 22$K 
in Y$_2$Mo$_2$O$_7$, we have measured the nonlinear
susceptibility coefficient, $\chi_3(T)$, which is expected to show a
power-law critical divergence at $T_f$~\cite{BYFH}:
\begin{equation}
\chi_3  \propto  \tau^{-\gamma} \;\;\;\;, 
\label{chi3-div}
\end{equation}
\noindent with $\tau\equiv T/T_f-1$ and $\gamma>0$. 
$\chi_3$ is extracted from the
temperature, $T$, and field, $H$, dependence of the magnetization,
$M(T,H) = \chi_1(T) H  - \chi_3(T)H^3 +
\chi_5(T)H^5 - ...$, 
where $\chi_1(T)$ is the linear susceptibility.  Hence, the
temperature dependence of $\chi_3(T)$ allows a determination of $T_f$
and $\gamma$~\cite{BYFH}.  In fact, all the nonlinear terms
$\chi_{2n+1}$ with $n\ge 1$ must diverge at $T_f$, since both $M(T,H$)
and $H$ are finite quantities.  It is therefore convenient to define a
net nonlinear susceptibility, $\chi_{\rm nl}$, as 
\begin{equation}
\chi_{\rm nl}(T,H) \equiv 1 - \frac{ M(T,H) }{\chi_1 H}  \;\;\;\;.
\label{nonlinsus} 
\end{equation} 
\noindent Right at $T_f$, $\chi_{\rm
nl}$ has a power law dependence on $H$:  
\begin{equation} 
\chi_{\rm
nl}(T=T_f,H) \sim H^{2/\delta}        \;\;\;\;, 
\label{nonlinsus-Tg}
\end{equation} 
\noindent where $\delta$ is a second {\it independent}
static critical exponent characterizing the spin-freezing
transition~\cite{BYFH}.  Finally, and perhaps the most relevant test
ascribing critical behavior to the spin-freezing transition is
obtained by seeking a scaling behavior of $\chi_{\rm nl}$ 
of the form
\begin{equation}
\chi_{\rm nl}(T,H) \propto H^{2/\delta}{\cal F}(\tau^{(\gamma+\beta)/2}/H)
 \;\;\;\; ,
\label{nonlinsus-scaling}
\end{equation} 
\noindent where $\beta$ is 
the spin-glass order parameter critical exponent~\cite{BYFH}.
Here, ${\cal F}(x)$ is the scaling function which must obey the following asymptotic behavior
\begin{eqnarray}
{\cal F}(x) & = & {\rm constant},\;\;\;\;   x\rightarrow 0  \;\;\;\;, \\
{\cal F}(x) & \propto & x^{-2\gamma/(\gamma+\beta)}, 
\;\;\; x\rightarrow \infty \;\;\; 
\label{asymptos}
\end{eqnarray}
\noindent in
order that the scaling behavior has ``physical content''~\cite{martinez}, 
and that Eqs.~\ref{chi3-div} and ~\ref{nonlinsus-Tg}
are recovered~\cite{BYFH,martinez}, hence giving the scaling relation
\begin{equation}
\delta=1+\gamma/\beta           \;\;\;\;.
\label{delta-gam-beta}
\end{equation}

The magnetization data were obtained 
under field-cooling conditions. The  field $H$, in the range
$[100-7000]$Oe, was switched on at  high temperature (70K $\sim 3T_f$),
and kept constant during subsequent {\it slow cooling} at a rate $5$ mK/s
in a fixed field, and down to the temperature of interest. 
It took over an hour to go from 70K to any of the $(H-T)$ data
point presented in our paper. This means almost twelve hours
to get to the lowest temperature.
The reason we cooled down in {\it  fixed field} instead of working
along conventional isotherms is to eliminate any possibility
of a magnetic field hysteresis of the superconducting magnet
in the magnetometer affecting the results;
the magnetic field was  never decreased
during the whole experiment.
Because of the irreversible and
time-dependent nature of the system's response below $T_f$, only results
in the temperature range $T_f<T<3T_f$ are included.  Our results on the low
temperature dynamical relaxation of the magnetization in
Y$_2$Mo$_2$O$_7$ will be reported elsewhere.
Three consecutive cooling runs for fixed field were performed, with the
magnetization data averaged over the three runs.

Prior to doing any
analysis, we have to deal with the fact that the interactions do not
perfectly average to zero, as evidenced by $\chi_1(T)$
not having a simple $\chi_1\sim 1/T$ Curie law.  It has been 
argued~\cite{BYFH,corr_scaling} several times
that the {\it leading} (first-order)
corrections to scaling coming from this non-zero
averaging of the interactions can be eliminated by fitting $\chi(T,H)$
to powers of $a_{2n+1}(T)\chi_1 [\chi_1 H]^{2n}$ for $n\ge 0$,
instead of simply $\chi_{2n+1}H^{2n}$, and considering the critical
behavior of $a_{2n+1}(T)$ instead of $\chi_{2n+1}(T)$. 
For each temperature the field dependence of $\chi$ at small field was
fitted with $\chi= \chi_1 - \chi_3 H^2$, giving $a_3=\chi_3/\chi_1^3$,
and varying the upper limit of the field range to determine the limit
of validity of this restricted fit beyond which higher $\chi_{2n+1}$
($n>1$) corrections become significant. The quality of our
magnetization data did not allow us to determine the
$a_5=\chi_5/\chi_1^5$
coefficient with precision better than $[50-100]$\%, and thus $a_5$
data are not included here. We show in
Fig. (2) the net nonlinear susceptibility, $\chi_{\rm nl}(T,H)$, as
defined in Eq.~\ref{nonlinsus}, with $\chi_1$ extracted from the fit $M(T,H)/H =
\chi_1 - \chi_3 H^2$, as a function of $H^2$ for few temperatures above
$T_f$.  These results emphasize the large increase of the nonlinear
susceptibility upon approching $T_f$.  We also notice
that $\chi_{nl}$ is only linear in $H^2$ up to a maximum field
$H_+(T)$ whose value is rapidly moving to zero upon approching
$T_f$ due to the turning on of the $\chi_{2n+1}$ ($n>1$) corrections
which themselves diverge at $T_f$ as $\tau^{(\beta-
n[\gamma+\beta])}$~\cite{BYFH}.
\begin{figure}
\begin{center}
  {
  \begin{turn}{90}%
    {\epsfig{file=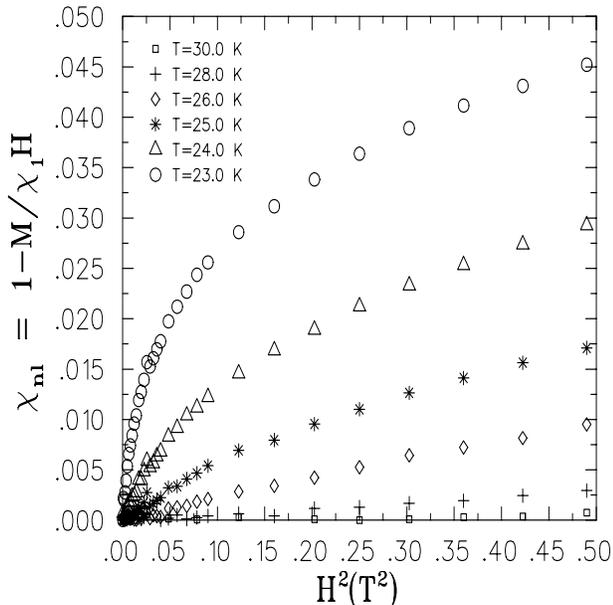,height=8cm,width=8cm} }
   \end{turn}
   }
\caption{Net nonlinear susceptibility,
$\chi_{\rm nl} = 1-M/\chi_1 H$ vs $H^2$ for the six temperatures
indicated.}
\end{center}
\end{figure}
Figure 3 shows a log-log plot of $\chi_3/\chi_1^3$
 vs $T/T_f-1$ for five different
choices of $T_f$.
The steepness of the curves increases as the chosen value
for $T_f$ is decreased.
Choosing $T_f>23.0$ would give us a curve with essentially no extended
range of power-law behavior ($\chi_3/\chi_1^3 \propto \tau^{-\gamma}$).
As can be seen in the top of Fig. 1, 
values of $T_f>23.0$K are definitely {\it above}
the FC-ZFC break-away point where the magnetization data acquisition runs 
are reversible at the lowest field.
Values of $T_f<21.0$ are also not possible to justify as they are clearly
well {\it below} the FC-ZFC break-away point.
Also, one observes an obvious upward curvature
at the far left of the data with the choice $T_f=21$K. 
Consequently, the log-log plot of the $\chi_3/\chi_1^3$ data strongly
suggests that $T_f$ is finite and between 
21.0K and 23.0K, and the extracted value of $\gamma$ depends on the choice of
$T_f$.  To better quantify this, we have fitted $\chi_3/\chi_1^3 \propto (1-T/T_f)^{-\gamma}$
for a fixed range of 
value $\chi_3/\chi_1^3 \in [2-30]$ for a {\it fixed} number of data
points (solid line fits on Fig. 3).
The goodness-of-fit, 
$X^2 = \sum_i [a_3^{\rm calc}(\tau_i) - a_3^{\rm meas.}(\tau_i)]^2$,
with $a_3  = \chi_3/\chi_1^3$ and $\tau_i = 1-T_i/T_f$ vs the choice of $T_f$ is
shown in the inset of Fig. 3, along with the corresponding variation of the
critical exponent $\gamma$. As can be seen, a pronounced mimimum in
$X^2$ is seen as a function of $T_f$, and the best fit 
is obtained for $T_f \approx 22.2K$, which
gives a value $\gamma \approx 2.8$.  Overall, a conservative estimate gives
$T_f\in [21.7 - 22.7]$K, giving $\gamma \in [2.4 - 3.4]$. 
\begin{figure}
\begin{center}
  {
  \begin{turn}{90}%
    {\epsfig{file=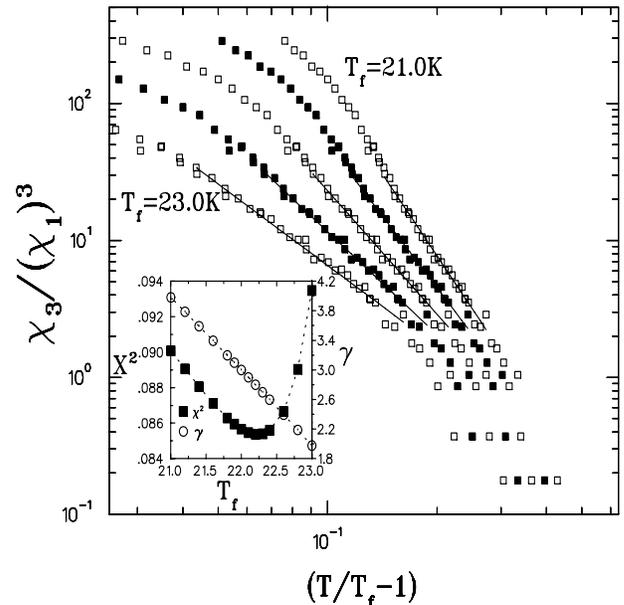,height=8cm,width=8cm} }
   \end{turn}
   }
\caption{Log-log plot showing the temperature dependence of 
$a_3=\chi_3/\chi_1^3$ vs $\tau\equiv T/T_f-1$ for five different
choices of $T_f$: $T_f=$ 21.0K, 21.5K, 22.0K, 22.5K, and 23.0K.
The inset shows the goodness of fit parameters, $X^2$, (filled
squares) for the solid line fits in the main panel, and
the exponent $\gamma$ (open circles) vs the chosen value for $T_f$
(see comments in text).}
\end{center}
\end{figure}
The power-law divergence of $a_3=\chi_3/\chi_1^3$
saturates for  $\tau < 0.05$ for a choice of $T_f\in [22.0-23.0]$K.
A reason for this levelling off of $a_3$ is that the range
of dominance of the term $\chi_3(T) H^2$ to $\chi_{\rm nl}$ falls
below the smallest field,  $H_{\rm min}$ ($H_{\rm min} \sim 100$Oe),
for which  good quality data
were obtained.  The increasingly important diverging higher
order terms of alternating signs ($\chi_5$, $\chi_7$, etc) contributing to
$\chi_{\rm nl}$  then cause
$a_3$ to be underestimated when
$H_+(T)$ becomes less than or equal to $\approx H_{\rm min}$.
Also, the slow but finite cooling rate inhibits
the correct equilibrium value of $\chi_{\rm nl}$ from being attained and this
effect may be compounded with the previous one to produce a saturation of
$a_3$ for $t<0.05$. Overall, the observed behavior  of $a_3$ for
Y$_2$Mo$_2$O$_7$ seen in Fig. 3, as well as the uncertainty on the value
of $\gamma$  are typical  
of  what is observed in conventional, chemically disordered  spin glasses.

We now attempt to verify that the spin-freezing in Y$_2$Mo$_2$O$_7$
is a legitimate critical phenomenon by seeking a data collapse and 
scaling behavior of
the net nonlinear susceptibility of
the form given in Eq.~\ref{nonlinsus-scaling}.
Choosing $T_f=22.2$K and $\gamma=2.8$, as found in the
inset of Fig. 3, we can find a reasonable data
collapse (scaling) of $\chi_{\rm nl}$ with a choice of $\beta=0.75 \pm 0.10$.
Had we worked with other choices of $T_f \in [21.7 - 22.7]$K, and
$\gamma \in [2.4 - 3.4]$, we would have found $\beta \in [0.6 - 0.9]$.
As found for $\gamma$ above, such a value for $\beta$ is typical
of that found in conventional disordered spin glases~\cite{BYFH}.
The scatter of the data at large $x$, 
$x \equiv \tau^{(\gamma+\beta)/2}/H$, arises because
these data are those that correspond to relatively high temperature 
(well above $T_f$) and small fields 
where the nonlinear magnetization 
is small and the experimental error is the largest.
In the limit $x\rightarrow 0$, 
we observe that ${\cal F}(x)$ approaches a constant value, hence confirming that
$\chi_{\rm nl} \propto H^{2/\delta}$ at $T=T_f$
with Eq.~\ref{delta-gam-beta} obeyed.
Taking  $\gamma=2.8$ and $\beta=0.75$ at $T=T_f=22.2$K, 
Eq.~\ref{delta-gam-beta} {\it predicts}
a value $\delta \approx 4.73$.  The inset of Fig. 4 shows a log-log
plot of $\chi_{\rm nl}(T=T_f=22.2){\rm K}$ vs $H$ with this value of $\delta=4.73$
(solid line). The fit is very good, confirming that the collapse of the 
$\chi_{\rm nl}(T,H)$ data has
physical meaning underlying a critical phenomenon which obeys
Eq.~\ref{delta-gam-beta}.
Also, in the limit of large $x$ (i.e. at small fields), we 
observe that the asymptotic behavior of ${\cal F}(x)$ is consistent with the power-law
${\cal F}(x) \propto x^{-2\gamma/(\gamma+\beta)}$ (dashed-line in main panel of Fig. 4) as,
again, it should be for the data collapse to have physical meaning and 
such that $\chi_{\rm nl} \sim \tau^{-\gamma} H^2$ for $T\gg T_f$ and
$H\rightarrow 0$. 
\begin{figure}
\begin{center}
  {
  \begin{turn}{90}%
    {\epsfig{file=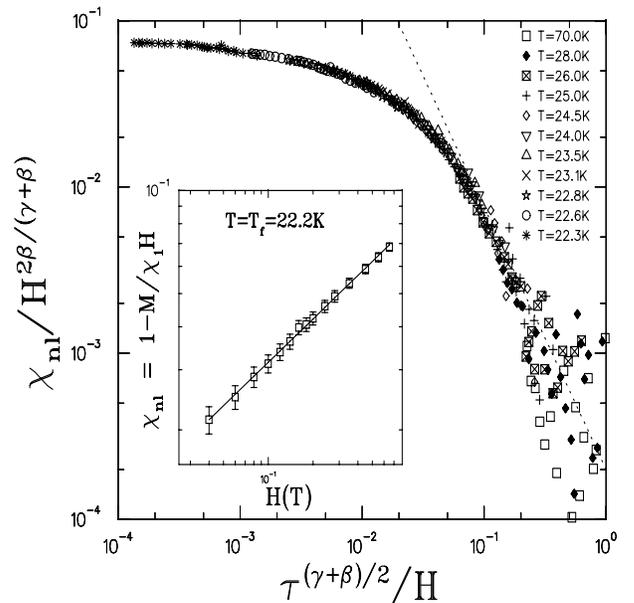,height=8cm,width=8cm} }
   \end{turn}
   }
\caption{Nonlinear magnetization analyzed according to a scaling model for
a nonzero spin-glass transition temperature with choices
$T_f=$22.2K, $\gamma=2.8$ and $\beta=0.75$.  The inset shows the log-log
plot of $\chi_{\rm nl}(T=T_f,H) \propto H^{2/\delta}$ with the value of
$\delta=4.73$ predicted from the scaling relation~\ref{delta-gam-beta},
with the values of $\gamma=2.8$ and $\beta=0.75$.}
\end{center}
\end{figure}
It is interesting to compare our results for the Y$_2$Mo$_2$O$_7$
pyrochlore with those for the SrCr$_8$Ga$_4$O$_{19}$ kagom\'e system
(SCGO)~\cite{martinez,ramirez}, and for the site-ordered gadolinium
gallium garnet magnet Gd$_3$Ga$_5$O$_{12}$ (GGG)~\cite{schiffer}.
Ramirez et al.~\cite{ramirez} found a power-law divergence of $\chi_3$
in SCGO with $\gamma \approx 2.4$, while Martinez et
al.~\cite{martinez} recently argued that
the freezing at $T_f \approx 3.5$K in SCGO is {\it not} associated with
a divergence of $\chi_3$ ($\chi_3$ was found to increase by a factor 5
or so in Ref.~\cite{martinez}), and that this material does not exhibit
conventional spin glass behavior.  Also, contrary to what we find for
Y$_2$Mo$_2$O$_7$, Martinez et al.  argued that the data-collapse (i.e.
scaling behavior) they obtained for the net nonlinear susceptibility of
SCGO was ``unphysical'' since the asymptotic behavior of their scaling
function was inconsistent with what it should have been according to the
scaling relation $\delta=1+\gamma/\beta$~\cite{martinez}.  Schiffer et
al.~\cite{schiffer} found a large increase of $\chi_3$ in GGG (6 orders
of magnitude between 0.2K and 5K), which they ascribe to a spin glass
transition. However, the temperature dependence of $\chi_3$ in GGG is
qualitatively different than what is found in conventional spin glasses
since $\chi_3$ has two maxima in GGG, while it is a monotonic function
of the temperature in conventional spin glasses as well as here in
Y$_2$Mo$_2$O$_7$.
Hence, from the point
of view of nonlinear susceptibility measurements, it therefore appears
that the spin glass behavior observed in Y$_2$Mo$_2$O$_7$ resembles
more closely what is found in conventional spin glasses than 
that which has been
found in other topologically frustrated antiferromagnets, such as SCGO
and GGG.

In conclusion, we have measured the nonlinear DC susceptibiltity of the
pyrochlore antiferromagnet Y$_2$Mo$_2$O$_7$ close to and above the
freezing temperature, $T_f \sim 22$K, where this material shows spin-glass behavior.
Our results  show that the freezing
transition observed in Y$_2$Mo$_2$O$_7$ is well characterized by a
power law divergence of the $\chi_3(T)$ nonlinear susceptibility
coefficient, $\chi_3(T) \propto (1-T/T_f)^{-\gamma}$, with a value
$\gamma \sim 3.0 \pm 0.5$.  This implies an  underlying thermodynamic glass
phase transition around 22K in this material.  This is further
supported by recent muon spin relaxation  measurements where a large
increase of the $1/T_1$ muon depolarization rate at $T\approx 20$K has
been observed, and which indicate a dramatic critical slowing down of
the Mo$^{4+}$ moments~\cite{uSR}.  The net nonlinear
susceptibility data, $\chi_{\rm nl}(T>T_f,H)$, can be collapsed onto a
scaling function from which we can extract the order pameter critical
exponent $\beta \sim 0.8 \pm 0.2$.  Right at $T_f$, we find
a behavior $\chi_{\rm nl} \sim H^{2/\delta}$ with a value of $\delta \sim
4.7$, which satisfies the scaling relation~\ref{delta-gam-beta}.
The values we find for $\gamma$ and $\beta$ are typical of what is
found in conventional chemically disordered spin glasses~\cite{BYFH}.

We hope that our results will stimulate further experimental and
theoretical studies
to clarify the origin of the spin-freezing transition and the
low-temperature glassiness seen in Y$_2$Mo$_2$O$_7$ 
and other geometrically frustrated magnetic  systems.

We thank D. Huse, A. Ramirez and 
G. Williams  for useful discussions, and M. Harris
for allowing us to use his postscript figure of
the pyrochlore lattice structure.  This research has been funded
by the NSERC under the NSERC Collaborative Research Grant {\it
Geometrically-Frustrated Magnetic Materials}.

\end{document}